\documentstyle{article}

\def\ni{\noindent }
\def\eq #1{(\ref{#1})}       
\def\u{u}                       
\def\yx{$\{x\leftrightarrow y\}$}

\newcommand{\be}[1]{\begin{equation}\label{#1}}
\def\ee{\end{equation}}

\newcommand{\ba}[1]{\begin{array}{#1}}
\def\ea{\end{array}}

\def\fr #1#2{\frac{#1}{#2}}

\def\se #1{sec.\,\ref{#1}}

\def\y1{\mbox{$y'$}}
\def\yt{\mbox{$y''$}}

\def\fz{f_0}
\def\fo{f_1}
\def\ft{f_2}
\def\f3{f_3}

\def\fl3{\tilde{\f3}}

\def\s3t{\tilde{{s_3}}}

\def\hyper3{{\bf hyper3}}   

\def\2F1{\mbox{$_2${F}$_1$}}
\def\1F1{\mbox{$_1${F}$_1$}}
\def\0F1{\mbox{$_0${F}$_1$}}
\def\PFQ{\mbox{$_p${F}$_q\;$}}

\begin{document}
\title{A connection between Abel and\\
\PFQ hypergeometric differential equations}

\author{E.S. Cheb-Terrab$^{a,b}$}

\date{}
\maketitle
\thispagestyle{empty}


\medskip
\centerline {\it $^a$CECM, Department of Mathematics}
\centerline {\it Simon Fraser University, Vancouver, British Columbia, Canada.}

\medskip
\centerline {\it $^b$Maplesoft, Waterloo Maple Inc.}


 \bigskip
 \centerline {\small{(Submitted to the {\it European Journal of Applied Mathematics} - February 2004)}}

\maketitle

\begin{abstract}

In a recent paper, a new 3-parameter class of Abel type equations, so-called
AIR, all of whose members can be mapped into Riccati equations, is shown.
Most of the Abel equations with solution presented in the literature belong
to the AIR class. Three canonical forms were shown to generate this class,
according to the roots of a cubic. In this paper, a connection between those
canonical forms and the differential equations for the hypergeometric
functions \2F1, \1F1 and \0F1 is unveiled. This connection provides a closed
form \PFQ solution for all Abel equations of the AIR class.

\end{abstract}

\section*{Introduction}

Abel-type ordinary differential equations (ODE), in either their second kind form \cite{kamke},

\be{abel_2k}
\y1 = \fr{\f3\, y^3 + \ft\, y^2 + \fo\, y + \fz}{g_1\,y + g_0},
\end{equation}

\ni where the $\{f_i,\ g_i\}$ are arbitrary functions of $x$, or their first
kind form obtained taking $\{g_1=0,\,g_0=1\}$, appear frequently in physical
applications \cite{kamke,green,sachdev}. This has for a long time motivated
their study. 

After pioneering work on their solutions by Abel \cite{abel}, the basis of
today's solving approach for these equations was set by Liouville
\cite{liouville2,liouville3} and Appell \cite{appell}, using classical
invariant theory. In brief, through transformations of the form

\be{tr_2k}
\{x = F(t),\ \ y= \fr{P_1(t)\,  \u + Q_1(t)}{P_2(t)\,  \u + Q_2(t)}\},
\ee

\ni where $\{F,\,P_1,\,P_2,\,Q_1,\,Q_2\}$ are arbitrary analytic functions
restricted only by $P_1\,Q_2 - P_2\,Q_1 \neq 0$, one can define an {\em Abel
class} of equations, all of whose members can be mapped between themselves
by means of \eq{tr_2k}. Then, a given Abel equation can be tackled by
formulating an {equivalence problem} between itself and a representative
of that class, for instance one whose solution is known.

A key ingredient in such an approach is the number of Abel classes that
possess a member - herein called ``solvable equation" - whose solution is
known. Through the class transformations \eq{tr_2k}, solvable equations
generate ``solvable classes". In a recent work \cite{abel1} (2000), it has
been shown that a large number of solvable Abel equations scattered in the
literature, including those presented by Abel, Liouville and Appell, are all
members of one or another of only four 1-parameter and seven 0-parameter
Abel solvable classes. In \cite{abel2} (2003), three new rather general Abel
classes were presented, so-called AIL, AIR and AIA, respectively depending
on 2, 3 and 4 parameters, and those solvable classes collected in
\cite{abel1} were in turn all shown to be particular cases of just these
three. Apart from the generalizing aspect of these new multi-parameter
classes, an important feature of AIL and AIR is that all their members can
respectively be mapped into first order linear and Riccati type equations;
that is, they are ``solvable" (AIL) or linearizable (AIR).

The presentation of these multi-parameter classes in \cite{abel2}, however,
didn't include a closed form solution for any member of the AIR
class. A solution was not known at that time. The authors acknowledged this
in \cite{abel2} and instead presented a partial classification of the members
of AIR as generated from three canonical forms, related to the roots of a
cubic. To know the closed form solutions for these canonical forms, however,
is important, because Riccati equations are not ``solvable" in general; only
with these solutions at hands can we think of the multi-parameter
AIR class as solvable.


In this paper, it is shown that a surprisingly simple connection exists
between the three AIR canonical forms and the differential equations
satisfied by the hypergeometric functions \2F1, \1F1 and \0F1. This connection
provides \PFQ closed form solutions for all members of the AIR
class, and permits formulating an alternative approach for resolving the
membership problem.

\section{The Abel Inverse Riccati - AIR - class}
\label{AIR_classification}

Recalling the material presented in \cite{abel2}, a multi-parameter class of
Abel equations, all of whose members can be transformed into Riccati
equations, can be obtained departing from the general form of a Riccati
equation,

\be{riccati}
\y1= f \left( x \right) {y}^{2}+g \left( x \right) y+h \left( x \right),
\end{equation}

\ni and applying to it the {\em inverse} transformation \yx\footnote{By \yx\ we
mean changing variables $\{x = u(t),\, y(x)=t\}$ followed by renaming
$\{u\rightarrow y,\, t\rightarrow x\}$.}, resulting in

\be{generic_inverse_riccati}
\y1 = \fr{1}{ f \left( y \right) {x}^{2}+g \left( y \right) x+h \left( 
y \right)}
\ee

\ni When all of $f$, $h$ and $g$ are of the form

\be{fgh}
y \rightarrow {\frac {s\,y + r}{a_{{0}} + a_{{1}}\,y + a_{{2}}\,{y}^{2} + a_{{3}}\,{y}^{3}}}
\ee

\ni where $\{a_i,\,s,\,r\}$ are constants with respect to $x$,
\eq{generic_inverse_riccati} has the form

\be{AIR}
\y1= {\frac {a_{{3}}\,y^3+a_{{2}}\,y^2+a_{{1}}\,{y}+a_{{0}}\,{y}}{ \left( s
_{{0}}+s_{{1}}\,x+s_{{2}}\,{x}^{2} \right) y+r_{{0}}+r_{{1}}\,x+r_{{2}}\,{x}^{
2}}}
\ee

\ni for some new constants $\{s_i,\,r_i\}$. This equation is of Abel
2nd kind type, can be transformed into a Riccati equation by means of \yx,
and is a representative of the most general Abel ODE class - so-called Abel
Inverse Riccati (AIR) - all of whose members are linearizable \cite{abel2}. In
\cite{abel2} it is shown that the 1-parameter classes presented by
Abel \cite{abel}, Liouville \cite{liouville3} and Appell \cite{appell}, as well
as most of the solvable examples found in the literature\footnote{For a
collection of these see \cite{abel1}.}, are all members of this AIR class,
generated from \eq{AIR} by applying the transformation \eq{tr_2k}. 

\subsection{Six-parameter AIR canonical forms}
\label{CanonicalForms}

An important property of \eq{AIR} is that its rational structure with
respect to $x$ and $y$ is invariant under M\"obius (linear fractional)
changes of $x$ and $y$. The problem of solving an arbitrary Abel equation
can then be formulated in two steps: bring the given equation to the AIR
form \eq{AIR} using the class transformations \eq{tr_2k}, then use M\"obius
transformations to reduce the AIR form to a canonical form we expect to be
able to solve.

With that in mind, in \cite{abel2}, M\"obius transformations of $y$ were
used to transform \eq{AIR} into three different canonical forms, and in
doing so, the number of parameters entering \eq{AIR} was reduced from ten to
six. No solutions to these canonical forms were known when
\cite{abel2} was written.


To construct these canonical forms of AIR with six parameters, following
\cite{abel2}, the numerator of \eq{AIR} is written in terms of its roots
$\rho_i$

\be{AIR_rho}
\y1={\frac{(y-\rho_0) (y-\rho_1) (y-\rho_2)}
{ \left( a_{{2}}\,x^2+a_{{1}}\,x+a_{{0}} \right) y + b_{{2}}\,x^2 + b_{{1}}\,x+b_{{0}}}}
\ee

\ni Then, using M\"obius transformations for the dependent variable
\begin{equation} y \rightarrow \fr{p + q\,y}{r + s\,y} \ee

\ni where $\{p,q,r,s\}$ are any constants satisfying $p\,s-r\,q \neq 0$,
\eq{AIR_rho} can be transformed into

\be{can_forms}
\y1={\frac {P \left( y \right) }{ \left( a_{{2}}\,x^2+a_{{1}}\,x+a_{{0}} \right) y + b_{{2}}\,x^2 + b_{{1}}\,x+b_{{0}}}}
\ee

\ni for some new constants $\{a_i\,b_i\}$, with $P(y)=1$, $P(y)=y$ or $P(y)
= y\,(y-1)$, respectively according to whether in \eq{AIR_rho} there are
only one, two or three distinct roots $\rho_i$. These three canonical forms
are shown in \cite{abel2} depending on six parameters, as in \eq{can_forms}.

\subsection{Three-parameter AIR canonical forms}
\label{CanonicalForms3}

Apart from the M\"obius transformations of $y$ used in \cite{abel2}, M\"obius
transformations of the independent variable,

\begin{equation}
x \rightarrow \fr{p+q\,x}{r + s\,x},
\ee

\ni also leave the structure of \eq{AIR} invariant, and so can be used to
further transform the three canonical forms represented by \eq{can_forms}.
In doing so, following \cite{theodore}, the number of parameters can be
reduced from six to three. For that purpose, in \eq{can_forms}, if any of
$\{a_1, a_2\}$ are different from zero, a transformation $x \rightarrow 1/x
+ \kappa$, with $\kappa$ being any non-zero root of
$a_{{2}}\,{\kappa}^{2}-a_{{1}}\,\kappa+a_{{0}} = 0$, will cancel the $a_2\,
x^2$ term in the denominator, taking the equation to a form with only five
parameters\footnote{If in \eq{can_forms} $a_1=0$ and $a_2=0$, then this
equation already depends on only four parameters.}. At this point, two
different cases can happen:

\medskip \ni \underline{Case A:} $a_1 = 0$, so applying the transformation
$x \rightarrow 1/x + \kappa$ just mentioned, \eq{can_forms} will be of the form

\be{can_forms3A}
\y1={\frac {P \left( y \right) }{ a_0\, y + b_{{2}}\,x^2 + b_{{1}}\,x+b_{{0}}}}
\ee

\ni for some new constants $\{a_0,\,b_i\}$.

\medskip \ni \underline{Case B:} $a_0 = 0$, hence applying that same
transformation $x \rightarrow 1/x + \kappa$, \eq{can_forms} will be of the
form

\be{can_forms3B}
\y1={\frac {P \left( y \right) }{ a_1\,x\, y + b_{{2}}\,x^2 + b_{{1}}\,x+b_{{0}}}}
\ee

\ni for some new constants $\{a_1\,b_i\}$. When both $a_1 \neq 0$ and $a_0
\neq 0$, a transformation of the form $x \rightarrow x - a_0/a_1$ will
transform \eq{can_forms} into an equation of the form \eq{can_forms3B}.

Summarizing, in all cases \eq{can_forms} can be transformed into an equation
of the form \eq{can_forms3A} or \eq{can_forms3B}, with only four parameters.

Finally, scaling $x \rightarrow \kappa\,x$ with $\kappa^2 = 1/b_2$ maps
both cases \eq{can_forms3A} and \eq{can_forms3B} into equations depending on
only three parameters\footnote{If $b_2=0$, \eq{can_forms3A} and
\eq{can_forms3B} depend on only three parameters; applying \yx\ one directly
obtains a first order linear ODE.}, of the form

\be{can_forms3}
\y1 = \fr{P(y)}{\tilde{a}\, y + (x-b) (x-c)}
\ee

\ni where $b$ and $c$ are constants, and in case A: $\tilde{a}$ is a
constant, while in case B: $\tilde{a} = a\,x$, and $a$ is a constant.

\subsection{Three, two, one and zero-parameter AIR canonical forms}
\label{CanonicalForms321}

Following \cite{abel2} and \cite{theodore} we have arrived at the three
canonical forms of AIR implicit in \eq{can_forms3}, according to whether
$P(y) = y\,(y-1)$, $P(y)=y$ or $P(y)=1$. Each of these canonical forms
splits into two cases, according to whether or not $\tilde{a}$ is constant,
and each of these six equations seem to depend on three parameters
$\{a,b,c\}$. Although that is the case when $P(y) = y\,(y-1)$ and
$\tilde{a}$ is non-constant, we will show below that the cases when
$\tilde{a}$ is constant or $P(y)=y$, and the case $P(y)=1$, respectively
depend on only two and one parameters at most.

Establishing this fact is of relevance since the classification of second
order \PFQ hypergeometric equations also splits into three canonical forms,
which are the equations satisfied by the \2F1, \1F1 and \0F1 functions, and these equations
respectively depend on three, two and one parameters. After the number
of parameters of the AIR canonical forms is precisely determined, one can
see that these classifications of Abel and \PFQ equations are
connected; this connection is discussed in the next section.

\subsubsection*{The parameters when the three roots $\rho_i$ in \eq{AIR_rho} are different}

Considering first the case where, in \eq{can_forms3}, $P(y) = y\,(y-1)$, if
$\tilde{a} \equiv a\,x$, \eq{can_forms3} depends on three parameters. If
$\tilde{a} \equiv a$ is a constant, however, a translation $x \rightarrow
x+b$ followed by renaming $c \rightarrow c+b$ transforms \eq{can_forms3}
into the representative of a 2-parameter class

\be{can_form_2F1b}
\y1={\frac {y \left( y-1 \right) }{a\,y+x \left( x-c \right) }}
\ee

\subsubsection*{The parameters when only two roots $\rho_i$ in \eq{AIR_rho} are different}

This case corresponds to \eq{can_forms3} at $P(y) = y$. By scaling $y
\rightarrow \kappa\, y$, the equation transforms into

\be{can_form_1F1}
\y1 = \fr{y}{\tilde{a}\,\kappa\, y + (x-b) (x-c)}
\ee

\ni Choosing $\kappa = 1/a$, the factor $\tilde{a}\,\kappa$ is either equal
to $1$ or to $x$ and in this way equation \eq{can_forms3} is reduced to an
equation representative of a 2-parameter class. 
When, in \eq{can_form_1F1}, $\tilde{a}\,\kappa = 1$, a translation $x
\rightarrow x + c$, followed by $c \rightarrow c+b$, transforms the equation into

\be{can_form_0F1}
\y1 = \fr{y}{y + x\,(x-c)},
\ee


\ni This is a canonical form representative of a 1-parameter class, which
happens to be equivalent to the 1-parameter Abel class presented by
Liouville in \cite{liouville3}, shown in \cite{abel1} as equation number
37.

\subsubsection*{The parameters when the three roots $\rho_i$ in \eq{AIR_rho} are equal}

This is equation \eq{can_forms3} at $P(y) = 1$; starting with the
case of non-constant $\tilde{a} \equiv a\,x$,

$$
\y1 = \fr{1}{a\,x\, y + (x-b) (x-c)},
$$

\ni using $y \rightarrow y/a + (b+c)/a$, this equation is transformed into

$$
\y1={\frac {a}{x\,y + {x}^{2} + b\, c}}
$$

\ni Scaling the variables $ \left\{ x \rightarrow x \sqrt
{a},\, y \rightarrow y \sqrt {a} \right\} $ further transforms the equation into

\be{can_form_1F1_1}
\y1={\frac {a}{a\,x\,y + a\,{x}^{2} + b\,c}}
\ee


\ni If $c \neq 0$, redefining $b \rightarrow
a\,b/c$, \eq{can_form_1F1_1} is reduced to the representative of a
1-parameter class

\be{can_form_1p_c}
\y1= \fr{1}{ x\,y + {x}^{2} + b}
\ee

\ni This equation happens to be equivalent to the 1-parameter Abel class
presented by Appell in \cite{appell} and shown in \cite{abel1} as equation
number 58. If, in \eq{can_form_1F1_1}, $c=0$, then $a$ is a factor which
cancels, leading to the representative of a 0-parameter class, 

\be{ode_liouville}
\y1= \fr{1}{ y + {x}^{2} },
\ee

\ni equivalent to one presented by Liouville in \cite{liouville3}, shown in
\cite{abel1} as equation number 35. 

The last case to consider occurs when, in \eq{can_forms3} at $P(y)=1$,
$\tilde{a} \equiv a$ is a constant:

\be{can_form_Bessel_particular}
\y1 = \fr{1}{a\, y + (x-b) (x-c)}
\ee

\ni If $b+c \neq 0$, using $\{x \rightarrow -\left( b+c \right) x,\, y
\rightarrow -{{y }/(b+c)} - {{b\,c}/{a}} \}$ followed by redefining $a
\rightarrow -a\,(b+c)^3$, this equation transforms into

\begin{displaymath}
\y1= \fr{1}{a\,y+{x}^{2}+x}
\end{displaymath}

\ni A further M\"obius transformation $ \{x \rightarrow x\sqrt [3]{a}-1/2,\
y \rightarrow {1/(4\,{a}) + {y}/{\sqrt [3]{a}}}\} $ takes this equation
into \eq{ode_liouville}. If, in \eq{can_form_Bessel_particular},
$b+c = 0$, using $\{x \rightarrow a^{1/2}\,x,\, y \rightarrow y/a^{1/2} +
b^2/a\}$ followed by redefining $a \rightarrow 1/a^2$, equation
\eq{can_form_Bessel_particular} is transformed into

\begin{displaymath}
\y1 = \fr{1}{a\,y + x^2}
\end{displaymath}

\ni A further scaling $\{ x \rightarrow t\sqrt [3]{a},\ y \rightarrow
y/\sqrt [3]{a} \}$ takes this equation also into \eq{ode_liouville}.


\subsection{Classification summary}
\label{ClassificationSummary}

Using M\"obius transformations of $y$ and $x$, \eq{AIR} is reduced to:

\begin{enumerate}

\item The representative of a 3-parameter AIR subclass, equation \eq{can_forms3} at
$\{P(y)=y\,(y-1),\, \tilde{a}\equiv a\,x\}$. This case occurs when the three
roots $\rho_i$ entering \eq{AIR_rho} are different.

\item The representative of a 2-parameter AIR subclass, equation
\eq{can_form_2F1b}, occurring when the three roots $\rho_i$ entering
\eq{AIR_rho} are different and in \eq{can_forms3} $\tilde{a}\equiv a$ is
constant.

\item The representative of a 2-parameter AIR subclass, equation
\eq{can_form_1F1} at $\kappa = 1/a$, corresponding to the case where only two of the roots $\rho_i$
entering \eq{AIR_rho} are different and in \eq{can_forms3} $\tilde{a}\equiv
a\,x$.

\item The representative of a 1-parameter AIR subclass, equation
\eq{can_form_0F1}, corresponding to the case where only two of the roots
$\rho_i$ entering \eq{AIR_rho} are different and in \eq{can_forms3}
$\tilde{a}\equiv a$ is constant.

\item The representative of a 1-parameter AIR subclass, equation
\eq{can_form_1p_c}, corresponding to the case where the three roots
$\rho_i$ entering \eq{AIR_rho} are equal and in \eq{can_forms3}
$\tilde{a}\equiv a\,x$.

\item The representative of a 0-parameter AIR subclasse, equation
\eq{ode_liouville}, happening when the three roots $\rho_i$ entering
\eq{AIR_rho} are equal and in \eq{can_forms3} $\tilde{a} \equiv a$ is constant.

\end{enumerate}

\ni M\"obius transformations map members of each of these six classes into
members of the same class, and the class representatives depending on one or
zero parameters (cases 4, 5 and 6) are equivalent to equations whose
solutions were presented by Liouville \cite{liouville3} and Appell
\cite{appell}.

\section{The connection with \PFQ second order hypergeometric equations}
\label{connection_to_pFq}

It is remarkable that the classification of the AIR class can be done in terms
of the multiplicity of the roots of a cubic and resulting in canonical forms
depending on three or less class parameters. A similar
classification is used for second order linear equations admitting \PFQ
hypergeometric solutions \cite{hyper3}. In that case, equations with three
regular singular points admit \2F1 hypergeometric solutions, depending on
three class parameters, and equations with one regular and one irregular
singular points admit \1F1 hypergeometric solutions, depending on two class
parameters. A special case of the latter happens when the parameters
entering $\1F1(a;\,b;\,x)$ are related by $a=b/2$, in which case \1F1 can be
re-expressed in terms of the $\0F1$ function through a quadratic
transformation. That case represents a third \PFQ class depending on only one
parameter.

A connection between this \PFQ classification and that for the Abel AIR equation
summarized in \se{ClassificationSummary} exists. This connection provides \2F1 solutions
when the three roots $\rho_i$ of the cubic in \eq{AIR_rho} are different
(cases 1. and 2. of \se{ClassificationSummary}), \1F1 solutions when only
two roots $\rho_i$ in \eq{AIR_rho} are different (cases 3. and 4. in
\se{ClassificationSummary}), and \1F1 or \0F1 solutions when the three roots
$\rho_i$ in \eq{AIR_rho} are equal (case 5 and 6 in
\se{ClassificationSummary}). The formulas relating these canonical forms of
AIR to the three \PFQ equations, and so relating the corresponding
solutions, can be derived as follows. 

\subsection{\2F1 solutions for the AIR equation \eq{AIR_rho} having three different roots $\rho_i$}

The Gauss or \2F1 hypergeometric linear equation,

\be{ODE_2F1}
\yt={\fr{(\alpha+\beta+1)\, x-\gamma}{x\,(1-x)}}\,\y1
+{\frac {\alpha\,\beta}{x\,(1-x)}}\,y,
\ee

\ni has for solution

\be{ans_2F1}
y = \2F1(\alpha,\beta;\,\gamma;\,x)\, C_1
+ {x}^{1-\gamma}\2F1(\alpha-\gamma+1, \beta-\gamma+1;\,2-\gamma;\,x)\, C_2
\ee

\ni where $C_1$ and $C_2$ are arbitrary constants. As with all second order
linear equations \cite{kamke}, using the change of variables $y \rightarrow
\exp(-\int \!{{y}/{x}}\,{dx})$, \eq{ODE_2F1} can be transformed into a
Riccati type equation

\be{ODE_R2F1}
\y1={\frac {{y}^{2}}{x}}
+ {\fr{\left( (\alpha+\beta)\,x-\gamma+1 \right)}{x \,(1-x) }}\,y
- {\frac {\alpha\,\beta}{1-x}}
\ee

\ni Applying the \yx\ transformation, the following Abel equation results

\be{ODE_A2F1}
\y1={\frac {y \left( y-1 \right) }{ \left( {x}^{2}- \left( \alpha+
\beta \right) x+\alpha\,\beta \right) y-{x}^{2}+ \left( \gamma-1 \right) x}}
\ee

\ni This equation is of the canonical form \eq{can_forms}, with $P(y)=y\,(y-
1)$, so it corresponds to the case where the three roots $\rho_i$ in
\eq{AIR_rho} are different. Using M\"obius transformations of $x$, equations
of this form can be transformed into the 3-parameter canonical form
\eq{can_forms3} of AIR as shown in \se{CanonicalForms3}. The cases
$\tilde{a} \equiv a\,x$ and $\tilde{a} \equiv a$ of \eq{can_forms3} are both
included in \eq{ODE_A2F1}, the latter corresponding to $\alpha+\beta = 0$,
and so resulting in an equation depending only on two parameters. These are
the cases 1. and 2. of the classification section \ref{ClassificationSummary}.

By construction, the solution to \eq{ODE_A2F1} is obtained changing \yx\
in the solution of \eq{ODE_R2F1}, which, in turn, is equal to $-\y1\,x/y$ with $y$ given in \eq{ans_2F1}. 

\subsection{\1F1 solutions for the AIR equation \eq{AIR_rho} having two
different roots $\rho_i$}

The confluent \1F1 hypergeometric equation,

\be{ODE_1F1}
\yt={\frac { \left( x-\beta \right)}{x}}\,\y1 + {\frac {\alpha}{x}}\,y,
\ee

\ni has for solution

\be{ans_1F1}
y = C_1\,{\rm M}(\alpha,\beta,x) + C_2\,{\rm U}(\alpha,\beta,x)
\ee

\ni where $C_1$ and $C_2$ are arbitrary constants and $M$ and $U$ are the
Kummer functions\footnote{${\rm M}(\alpha,\beta,x) = \1F1(\alpha;\,\beta;\,x)$.}
\cite{abramowitz}. Using the same two changes of variables which transform
the \2F1 equation \eq{ODE_2F1} into the Abel form \eq{ODE_A2F1}, equation
\eq{ODE_1F1} is transformed into the Abel equation

\be{ODE_A1F1}
\y1={\frac {y}{ (x-\alpha)\, y + x\,(x-\beta+1)}}
\ee

\ni which is of the canonical form \eq{can_forms}, with $P(y) = y$, and so
it corresponds to the case where two of the three roots $\rho_i$ in
\eq{AIR_rho} are equal. As shown in the previous section, equations of this
form can be reduced to the 2-parameter canonical form \eq{can_form_1F1} of
AIR. Concretely, changing $x \rightarrow x+\alpha$ followed by renaming
$\beta \rightarrow \beta + \alpha + 1$ and $\alpha \rightarrow -\alpha$,
\eq{ODE_A1F1} transforms into

\be{oc_11}
\y1={\frac {y}{x\,y+ \left( x-\alpha \right)  \left( x-\beta \right) }}
\ee

\ni This equation has the form \eq{can_form_1F1} for $\tilde{a}\,\kappa =
x$, hence this is case 3 of the classification section
\ref{ClassificationSummary}. A solution to this canonical form \eq{oc_11},

   
\begin{equation}
C_1+{\frac {x\,{\rm M}( -\beta,1+\alpha-\beta,y) -
\beta\,{\rm M}( 1-\beta,1+\alpha-\beta,y) }{x\,{\rm U}( -\beta,1+\alpha-\beta,y) 
+ \alpha\,\beta\,{\rm U}( 1-\beta,1+\alpha-\beta,y) }} = 0,
\ee

\ni is obtained
applying to the solution of \eq{ODE_A1F1} the same transformations used to map \eq{ODE_A1F1} into \eq{oc_11}.
A solution for \eq{ODE_A1F1} is obtained from the solution \eq{ans_1F1} as explained in the \2F1
case.

If, instead of departing from the general form \eq{ODE_1F1} for \1F1, one
departs from the particular \1F1 equation admitting \0F1 solutions,
that is,

\be{ode_0F1}
\yt=-{\frac {c}{x}}\, \y1 + {\frac {y}{x}},
\ee

\ni and applies the same two changes of variables used to transform the
\2F1 and \1F1 equations into Abel forms, followed by changing $y \rightarrow -y$ and
renaming $c \rightarrow c+1$, the resulting Abel equation is identical to
the AIR canonical form \eq{can_form_0F1}. So the \0F1 equation can be
associated to case 4 of \se{ClassificationSummary}. The solution to
\eq{can_form_0F1} can then be expressed directly in terms of the Bessel
functions of the first and second kind, ${\rm J_c}(x)$ and ${\rm Y_c}(x)$, as

\begin{equation}
C_1-{\frac {x\,{\rm J_c} \left(2\,\sqrt {y} \right) -{\rm J_{c+1}} \left(2\,\sqrt {y} \right) \sqrt {y}}
{-x\, {\rm Y_c} \left(2\,\sqrt {y} \right) + {\rm Y_{c+1}} \left(2\,\sqrt {y} \right) 
\sqrt {y}}}=0
\ee


\subsection{\1F1 and \0F1 solutions for the AIR equation \eq{AIR_rho} having
three equal roots $\rho_i$}

Equation \eq{can_form_1p_c}, presented as case 5 in
\se{ClassificationSummary}, is the 1-parameter canonical form of AIR
equivalent to the one presented by Appell in \cite{appell}. Its connection
with the \1F1 equation can be derived as follows. Applying \yx\ to the AIR
form \eq{can_form_1p_c}, in order to obtain its Riccati form, then changing
$y \rightarrow -\y1/y$, we obtain the second order linear form

\be{draft_22}
\yt=x\,\y1-b\,y
\ee

\ni This equation, in turn, can be obtained from the \1F1 equation \eq{ODE_1F1} by
changing $\{x \rightarrow x^2/2,\ y \rightarrow y/x\}$ and
evaluating its parameters at $\{\alpha = 1/2-b/2,\ \beta = 3/2\}$. So
applying the same transformation to the solution \eq{ans_1F1} of the \1F1
equation we obtain the solution to \eq{draft_22}; applying to this result
the inverse of the transformations used to map \eq{can_form_1p_c} into
\eq{draft_22}, we obtain the solution of the AIR form \eq{can_form_1p_c} as


\begin{equation}
C_1
+ {\fr 
    {2\, (1-b) {\rm M}( (3-b)/2,\, 3/2,\, {y}^{2}/2) + 2\,{\rm M}((1-b)/2,\, 3/2,\, {y}^{2}/2)\, ( b + x\,y) }
    {b \,(b-1) {\rm U}( (3-b)/2,\, 3/2,\, {y}^{2}/2) + 2\,{\rm U}((1-b)/2,\, 3/2,\, {y}^{2}/2)\, ( b + x\,y) }
    }
\ee


Finally, for the 0-parameter AIR canonical form \eq{ode_liouville}, that is,
case 6 of the summary in \se{ClassificationSummary}, a solution can be
obtained as in the previous case, applying to \eq{ode_liouville} \yx, followed by $y
\rightarrow -\y1/y$, resulting in its second order linear form. Resolving an
equivalence between this linear form and the \0F1 equation \eq{ode_0F1},
then reversing the transformations used, a solution for \eq{ode_liouville},
in terms of the Airy functions\footnote{ ${{\rm Ai}(z)}={\fr {{\0F1(\
;\,2/3;\,1/9\,{z}^{3})}\,\sqrt [3]{3}} {3\,\Gamma \left( 2/3 \right) }}
-{\fr {z\ {\0F1(\ ;\,4/3;\,1/9\,{z}^{3})}\,\Gamma( 2/3)\,\sqrt [6]{3}}
{2\,\pi }}$ } ${\rm Ai}(x)$ and ${\rm Bi}(x)$, is

\begin{equation}
C_1 + {\frac {x\,{{\rm Bi}(-y)}-{{\rm Bi'}(-y)}}{x\,{{\rm Ai}(-y)}-{{\rm Ai'}(-y1)}}}=0
\ee

\section{Conclusion}

This paper presented a complete classification of the Abel Inverse Riccati
(AIR) class of equations, which is the most general Abel class all of whose members are
linearizable \cite{abel2}. With this classification at hands, a direct
relation between the canonical forms of AIR presented in
\se{AIR_classification} and the second order linear equations for the
hypergeometric functions \2F1, \1F1 and \0F1 was established in
\se{connection_to_pFq}.

The first important consequence of this connection is that it makes the
whole AIR class of Abel equations solvable: through the class
transformations \eq{tr_2k}, the connections to \PFQ hypergeometric equations provide a
\PFQ closed form solution to any Abel equation member of the
AIR class, as shown in \se{connection_to_pFq}.

This connection also permits tackling the membership problem with respect to
AIR using a different and simpler approach than the traditional one.
Let us recall that the traditional approach consists of formulating an
equivalence between a given Abel equation and each of the canonical forms of
AIR. When the equivalence is possible, this approach also requires
computing the values of the class parameters for which the equivalence
exists. Even with the powerful computers currently available and using the
most modern symbolic algebra packages, such an approach is unrealistic: when
the number of class parameters is greater than one, the computation involves
composed multivariable resultants, resulting in untractable expression swell
\cite{abel1}.

An alternative approach, exploring the results of \se{AIR_classification},
consists of splitting the equivalence process into two steps. In the first
step, one attempts an equivalence to the AIR ``form" \eq{AIR} (symmetry and
integrating factor techniques can be of use for this purpose), not requiring
the computation of the value of the class parameters. In a second step, one
formulates the tractable and relatively easy problem of an equivalence under
M\"obius transformations of $x$ and $y$, between the AIR form \eq{AIR}
obtained for the given equation and each of the canonical forms of AIR
summarized in \se{ClassificationSummary}. This second step leads to the
values of the class parameters resolving the equivalence and in that way to
a solution for the problem. An implementation of these ideas using symbolic
algebra software is currently under development.

Finally, the connection with \PFQ linear equations shown in
\se{connection_to_pFq} indicates that other connections between
\mbox{$_p${F}$_q$}, Elliptic and Heun type functions is possible; work on
this topic is in progress.

\medskip
\ni {\bf Acknowledgments}
\medskip

\noindent This work was developed in the framework of the MITACS project and
supported by both the Centre of Experimental and Constructive Mathematics,
Simon Fraser University and the Maplesoft division of
Waterloo Maple Inc. The author would like to thank K. von B\"ulow for a
careful reading of this paper, and T. Kolokolnikov and A.D. Roche for
fruitful discussions.


\begin{thebibliography}{99}

\bibitem{kamke} E. Kamke, {``Differentialgleichungen"}, N.Y. Chelsea Publ. Co. (1947).

\bibitem{green} {A.D. Polyanin, V.F. Zaitsev},
``Handbook of Exact Solutions for Ordinary Differential Equations".
CRC Press, Boca Raton (1995).

\bibitem{sachdev} P.L. Sachdev, ``A Compendium of Nonlinear Ordinary
Differential Equations". John Wiley \& Sons (1997).

\bibitem{abel} N.H. Abel, ``Oeuvres Compl\'etes II". S.Lie and L.Sylow,
Eds., Christiana, 1881.

\bibitem{liouville2} R. Liouville, ``Sur une classe d'\'equations
diff\'erentielles du premier ordre et sur les formations invariantes
qui s'y rapportent", Comptes Rendus {\bf 103}, 460-463 (1887).

\bibitem{liouville3} R. Liouville, ``Sur une \'equation diff\'erentielle du
premier ordre", Acta Mathematica {\bf 26}, 55-78 (1902). See also R.
Liouville, Comptes Rendus {\bf 103}, 476-479 (1886).

\bibitem{appell} P. Appell, ``Sur les invariants de quelques \'equations diff\'erentielles",
Journal de Math\'ematique {\bf 5}, 361-423 (1889).

\bibitem{abel1} E.S. Cheb-Terrab, A.D. Roche, ``Abel Equations: Equivalence
and Integrable Classes", Computer Physics Communications 130, 197 (2000).

\bibitem{abel2} E.S. Cheb-Terrab, A.D. Roche, {``An Abel ODE class generalizing
known integrable classes"}, European Journal of Applied Mathematics, Vol.
14, No. 2, pp. 217-229 (2003).

\bibitem{theodore} T. Kolokolnikov, private communication.

\bibitem{hyper3} L.Chan, E.S. Cheb-Terrab, ``Non Liouvillian solutions for
second order linear ODEs", submitted for ISSAC'04, Spain (Jan/2004).

\bibitem{abramowitz} M. Abramowitz and I. A. Stegun, {``Handbook of
mathematical functions"}, Dover (1964).


\end{thebibliography}
\end{document}